# Identification of the most important external features of highly cited scholarly papers through 3 (i.e., Ridge, Lasso, and Boruta) feature selection data mining methods

**Sepideh Fahimifar[1],**
**Khadijeh Mousavi[2],**
**Fatemeh Mozaffari[3],**
**Marcel Ausloos[4, 5, 6,*]**

[1]Department of Information Science and Knowledge Studies, University of Tehran, Tehran, Iran. Email: sfahimifar@ut.ac.ir ORCID: 0000-0001-5182-9159

[2]Faculty of Management, University of Tehran, Tehran, Iran. Email: mousavi.khadijeh@ut.ac.ir

[3]Department of Information Technology Management, University of Tehran, Tehran, Iran. Email: famozaffari@ut.ac.ir

[4]School of Business, University of Leicester, Brookfield, Leicester, LE2 1RQ, United Kingdom. Email: ma683@le.ac.uk

[5]Department of Statistics and Econometrics, Bucharest University of Economic Studies, Calea Dorobantilor 15-17, Bucharest, 010552 Sector 1, Romania. Email: marcel.ausloos@ase.ro

[6]GRAPES, rue de la Belle Jardiniere, 483/0021, B-4031, Liege Angleur, Belgium. marcel.ausloos@uliege.be ORCID: 0000-0001-9973-0019

- Corresponding Author:

Prof. M. Ausloos, School of Business, University of Leicester, Brookfield, Leicester, LE2 1RQ, UK

Telephone number: +3243714340

**Abstract**

Highly cited papers are influenced by external factors that are not directly related to the documents intrinsic quality. In this study, 50 characteristics for measuring the performance of 68 highly cited papers, from the Journal of The American Medical Informatics Association indexed in Web of Science (WoS) from 2009 to 2019 were investigated.

In the first step, a Pearson correlation analysis is performed to eliminate variables with zero or weak correlation with the target ("dependent") variable (number of citations in WoS).

Consequently, 32 variables are selected for the next step. By applying the Ridge technique, 13 features show a positive effect on the number of citations. Using three different algorithms, i.e., Ridge, Lasso, and Boruta, 6 factors appear to be the most relevant ones.

The "Number of citations by international researchers", "Journal self-citations in citing documents", and "Authors self-citations in citing documents", are recognized as the most important features by all three methods here used.

The "First author scientific age", "Open-access paper", and "Number of first author's citations in WOS" are identified as the important features of highly cited papers by only two methods, Ridge and Lasso. Notice that we use specific machine learning algorithms as feature selection methods (Ridge, Lasso, and Boruta) to identify the most important features of highly cited papers, tools that had not previously been used for this purpose. In conclusion, we re-emphasize the performance resulting from such algorithms. Moreover, we do not advise authors to seek to increase the citations of their articles by manipulating the identified performance features. Indeed, ethical rules regarding these characteristics must be strictly obeyed.

**Highlights:**

- Comparing 3 feature selection methods: Ridge, Lasso, Boruta
- Analyzed data set relies on a representative sample from a health informatics journal
- Highly cited papers are studied
- ''External'' features which affect a high citation level are addressed

# 1. Introduction

It is of common knowledge that it is quite difficult to find an appropriate way to quantify the quality of papers. Often, the number of scholars' publications and the number of citations they receive have both become the main criteria for their impact evaluation (Ale Ebrahim et al. 2013; Rhaiem 2017), - loosely identified as their quality. In particular, the citation number has long been the benchmark for evaluating and demonstrating the usefulness of the scientific work (Antoniou et al. 2015; Moral-Muñoz et al. 2015; Kolle et al. 2016; Noorhidawati et al. 2017; Aksnes et al. 2019). This usefulness comes out from the sources, ideas, and results of articles which are identified as of interest and subsequently cited by other researchers (Zhang & Guan 2017). Therefore, articles with a very high number of citations appear to be scientifically more valuable because they are apparently influential on the work of many other researchers (González-Betancor & Dorta-González 2017), thereby appearing to have a special impact on the advancement of science (Bauer at al. 2016). The basic assumption is that, on the same subject and approximately at the same time of publication, those papers receiving many citations are more qualified than those receiving few citations (Bornmann et al. 2012). Therefore, by an audacious extension, highly cited papers and lowly cited papers are considered to represent high-quality and low-quality scientific papers, respectively (Kostoff 2007). Accordingly, due to the likely high quality of highly cited articles, these can be used as a criterion for evaluating the performance of an organization or of a scientist and consequently play a role in funding policies, promotions, etc. (Wang et al. 2019).

Moreover, these highly cited papers are much more visible and can make more famous the authors organizational affiliation (Antonakis et al. 2014; Gutiérrez-Salcedo et al. 2018). Martínez et al. (2015) also consider that "*highly cited papers of a scientific discipline help to discover potentially important information for the development of a discipline and understand the past, present, and future of its scientific structure*". This can also help in understanding the structuring of research fields (Song & Kim 2013). In addition, the presence of several highly cited papers in a journal can increase the impact factor of the journal (Liu et al. 2017). Thus all such papers provide examples of the Matthew effect (Merton 1968).

However, various external factors that do not appear to be directly related to the article's content intrinsic "quality" do influence its number of citations (Van Wesel et al. 2014; Onodera & Yoshikane 2015; Letchford et al. 2015). For example, the high number of citations might result from a multiple set of biases rendering citation-based measures practically flawed (Moustafa 2016). A main cause is the biased citation procedure, with voluntarily or not aspects, in which for various reasons one prefers to cite a paper in which a co-author is "famous" rather than a paper from a research group of a less famous institution. This results again into a Matthew effect (Merton 1968). Another cause resides in the necessary collaborative networks schemes increasing the number of co-authors per paper, whence inducing a large number of self-citations thereafter; these are particularly relevant in health innovation and global health reports (Fonseca et al. 2016).

A third cause can be the voluntary manipulation through self-citations or through citation cartels described as groups of authors that cite each other disproportionately more than they do other groups of authors that work on the same subject (Fister et al. 2016). Let us also mention that coercive citations (Wilhite & Fong 2012; Chang, McAleer, & Oxley 2013; Herteliu et al. 2017) have some biased impact as well.

In contrast, there can be a boycott of research groups and specific researchers in order to let them appear as of weak quality, - a relevant criterion at the competitive time of funding; this boycott can also be used in order to weaken the possible call of researchers as invited, - thus "glorified", speakers at scientific meetings. Other citation boycott causes can find roots in various scientific behavior aspects, like so called but undefined "personal preferences'' (Camacho-Miñano & Núñez-Nickel 2009; MacRoberts & MacRoberts 2010).

Thus, correctly, Haslam et al. (2008) pointed out that the criteria for determining the quality of a research work include subjective and objective criteria. The subjective criteria are based on the document's intrinsic attributes, which are often judged by an expert prior to publication. On the other hand, the objective criteria are based on the document's extrinsic attributes, - their effect being revealed after publication. There is a marked discrepancy obviously between these subjective and objective criteria; yet, if both are appropriately considered, they can increase the quality and impact of subsequent researches (Haslam et al. 2008), - in a quite positive way therefore.

Notice that somewhat amazingly there are various definitions of "highly cited papers" in the literature. According to Aksnes et al. (2003b), there are two approaches, including absolute thresholds or relative thresholds for defining a highly cited paper. When a fixed citation value is used, the difference between different subject areas is not considered; however, in the relative threshold approach, the subject area or the discipline also plays a role in determining whether an article is highly cited (Aksnes 2003b). For example, some studies have considered articles with at least 100 citations as a highly cited paper (Madhan et al. 2010; Wang et al. 2011; Ho 2012; Chen & Ho 2015; D. Ivanović & Ho 2016; Elango & Ho 2017; Kolle et al. 2017; Mo et al. 2018; X. Zhang et al. 2019; L. Ivanović & Ho 2019). However, others have extracted highly cited papers based on the *Essential Science Indicators* (ESI) definition (Miyairi & Chang 2012; Dorta-González & Santana-Jiménez 2019; Liao et al. 2019). According to the ESI[1] Web of Science Database, "Highly cited papers reflect the top 1% of papers by field and publication year".

Therefore, in the present study, in accordance with ESI definition, we have selected the highly cited papers of a particular journal, *Journal of the American Medical Informatics Association (JAMIA)*, as the study sample. We do not estimate that there is any strong assumption in so doing. The methodology[2] seems to us quite widely applicable. For completeness, notice that *JAMIA*, indexed in *Web of Science* (WOS), publishes articles in five categories according to *Journal Citation Reports* (JCR): Computer Science (Information Systems and Interdisciplinary Applications), Health Care Sciences & Services, Information Science & Library Science, Medical Informatics, which is Q1 in 3 domains and Q2 in 2 domains (as of July 12, 2021). *JAMIA* also ranks first in the field of libraries and information science, second in medical information, fourth in health care sciences & services, and sixth in computer science.

Within this framework, the present study examines the characteristics of (*JAMIA*) highly cited papers during a recent 10-year period, from 2009 to 2019, in order to contribute to an answer for the fundamental question: "What are the most important features of (these) highly cited papers which can affect the number of article's citations?"

---

[1] https://clarivate.com/webofsciencegroup/solutions/essential-science-indicators/

Our methodology goes as follows. Consider that feature selection is a multi-faceted and evolving problem; while univariate feature ranking with correlation coefficients is one of the simplest methods for feature selection, more complex methods can be used, especially when looking for causal feature selection (Guyon 2008). In our research, the correlation analysis is used as the initial selection of the variables. Next, we use three feature selection[2] techniques to identify the most important features of highly cited papers in a particular journal: the (i) Ridge, (ii) Lasso, and (iii) Boruta feature selection techniques. We propose this type of two-step approach in view of contrasting the methodology with others found through some literature review.

## 2. Literature Review. State of the art.

Many researchers have investigated the factors contributing to increase the number of citations of articles. Aversa (1985) seems one of the pioneers when illustrating citation patterns of highly cited papers. Findings of her research confirmed Price's studies on citation aging. In particular, Price (1965, 1976) had shown that the Number of highly cited papers (exponentially) decreases more slowly than less cited papers. Aksnes (2003b) identified the most important features of 297 highly cited papers published from 1981 to 1996, - which had at least one Norwegian author. Features such as publishing in a Journal with a high impact factor, Number of authors, Citations from non-Norwegian researchers, and International collaboration are found to be among the most effective features for getting a large citation number. According to the findings of his study, Self-citation has an only small share factor in the highly cited papers (Aksnes 2003b).

In the meantime, Kostoff (2007) compared the numerical characteristics (Numbers of authors, References, Citations, Abstract words, and Journal pages), organizational characteristics (First author's country, Institution type, and Institution name), and medical characteristics (Medical condition, Study approach, Study type, Study sample size, and Study outcome) of highly and poorly cited articles in the medical field, - papers published in The Lancet. Overall, the highly cited articles have more authors, more abstract lengths, more references, and more pages.

Haslam et al. (2008) have also categorized relevant factors into Author characteristics, Institutional factors (i.e., University prestige, Journal prestige, Grant support), Structure-related features (i.e., Title characteristics, Figures and tables, Number and Recency of references), and Research approach. In addition, Onodera & Yoshikane (2015) reviewing these findings, divided 15 contributing factors into five categories: Authors' degree of collaboration, Cited references, Article's visibility, Authors' past achievements, and Publishing journal. Tahamtan et al. (2016) considered 28 citation-related factors, divided into three categories: Article-related, Author-related, and Journal-related factors. Xie et al. (2019) examined 66 factors in four categories: Article-related, Author-related, Reference-related, and Citation-related factors. Stevens et al. (2019) investigated the impact of two Journal-related, two Author-related, one Citation-related, and 34 Article-related factors.

---

[2] The feature selection method is discussed in the methodology section of this article.

Moreover, other factors have been considered. Let us mention: Title characteristics (Jacques & Sebire 2010; Habibzadeh & Yadollahie 2010; Paiva et al. 2012), Number of pages, Number of words in title, Number of references, Sentences in the abstract, Sentences in the paper, Number of authors and Readability (Van Wesel et al. 2014), the Title (Letchford et al. 2015; Alimoradi et al. 2016) and other Morphological characteristics of articles (Alimoradi et al. 2016), Collaboration, i.e., number of authors or institutions (Figg et al. 2006), Impact of scientific affiliation and Intellectual base (Zhang & Guan 2017), Frequency of paper's keywords per journal and keyword repetition in the abstract with regard to Abstract length (Sohrabi & Iraj 2017), Keyword popularity (Hu et al. 2020), Manuscript length, Number of authors and Number of references (Fox et al. 2016), and also the effect of Wikipedia (Marashi et al. 2013).

Moreover, some scholars have investigated the characteristics of highly cited papers with a focus on a specialized area (Ho 2014; Chen & Ho 2015; Knudson 2015; D. Ivanović & Ho 2016; Khan et al. 2017; Kolle et al. 2017; Moral-Munoz et al. 2018; N. Zhang et al. 2018; Mo et al. 2018; L. Ivanović & Ho 2019; X. Zhang et al. 2019; Liao et al. 2019) or focusing on a particular country such as Korea (Krajna & Petrak 2019) or India (Elango & Ho 2017). Also, 1857 highly cited review with at least a thousand citations published between 1906 and 2010 were investigated by (Ho & Kahn 2014). According to most of these studies, the United States authors have often contributed the most to produce such articles (Chen & Ho 2015; D. Ivanović & Ho 2016; Kolle et al. 2017; N. Zhang et al. 2018; Mo et al. 2018; Liao et al. 2019; L. Ivanović & Ho 2019). Notice that the impact of factors of journals and papers in each mentioned study varies depending on the sample. Some of these most relevant studies are presented in detail next.

Elgendi (2019) examined 200 highly and lowly cited papers in order to extract their features and comparing them. The findings of Elgendi's research show that there is a significantly negative correlation between the Number of citations and the Paper title length; in contrast, the Number of views, tables, shapes, characters, and authors have a significantly positive correlation with the Number of citations. However, the Number of equations does not correlate with the Number of citations.

In a study by Dorta-González & Santana-Jiménez (2019), 10,000 scientific papers (indexes in the WOS database) from 2007 to 2016 were examined. These authors used the median as the measure of central and nonparametric median tests to compare papers, i.e., those highly cited and those not highly cited. Their findings show that factors such as having more Authors, Article pages, and Bibliographical references, in addition to publishing in journals with a high impact factor and having slightly shorter titles and longer abstract, could increase the Number of citations of one article. Their results show that there is no (linear) correlation between any other pair of variables analyzed so far.

An analysis based on descriptive observations rather than statistical ones was carried out by Miyairi & Chang (2012). Their study analyzed the bibliometric characteristics of 91,428 highly cited papers published between 2000 and 2009 in Taiwan extracted from the ESI Web of Science. The cooperation factor has much impact: in recent years, Taiwan has the highest international cooperation with countries like the United States, China, Germany, and Japan to produce highly cited papers in comparison to its neighboring countries in Asia.

Noorhidawati et al. (2017) conducted a research study on 708 Malaysian highly cited papers from 2006 to 2016 to identify the characteristics of Malaysian highly cited papers; they identified nine characteristics. The results of their research show that although the highly cited papers are more often an article than a review paper, the review papers have a higher citation impact than the articles. Malaysian highly cited papers are mainly published in high impact factor journals. Most of these papers have a large number of authors and are funded by national and international funders and corporative funding.

Thus, it can be observed that an overall consistent finding pertains to the impact factor of the journal where the highly cited paper is published. Indeed, according to many research results, a journal impact factor has a positive effect on increasing the number of citations of a research work (Vanclay 2013; Falagas et al. 2013; Khan et al. 2017; Stevens et al. 2019), as Judge et al. (2007) had already claimed: "the single most important factor driving citations to an article is the prestige or average citation rate of the journal in which the article was published".

Therefore, in order to ignore the effect of journal impact factor as some independent variable, masking other features, in the present study, we focus on highly cited papers from a particular journal. Moreover we use an original methodology. That means, in our study, we use Ridge, Lasso, and Boruta techniques to feature selection,- which has not been used so far (to the best of our knowledge) to identify the most important features of highly cited papers. Moreover, within this present framework methodology, one can interestingly notice that one can also try to predict the occurrence of highly cited papers through an early detection mechanism for candidate breakthroughs as discussed by Ponomarev et al. (2014).

## 3. Methodology

### a) Search Strategy

Data for the highly cited papers of the *Journal of The American Medical Informatics Association* on June 1, 2019, were extracted from the WOS database. For this purpose, the title of the journal was searched in the publication name field; the search results were limited to highly cited papers. WOS database covers journal citation data since 1975.

The present research has been conducted in a three steps methodology; *first step*: bibliographic study and identifying the potential factors affecting the number of citations, *second step*: feature extraction, and *third step*: feature selection and data analysis to identify the most effective features of the studied papers.

*First step*: Firstly, to identify important features concerning the number of citations, a bibliographic study has been conducted. In this phase, "highly cit* paper", "highly cit* article*", "citation impact", and "increases citation*" keywords have been used for a search in the title field by using OR operator. The first search result contained around 380 documents. Other documents were identified based on references to these 380 documents. Finally, about 50 potential factors were selected from reading 48 papers (Table 1).

### b) Feature Extraction

*Second step*: In the second step, 50 numerical and nominal factors (also called "features") were extracted, as presented in Table 1. Some of the factors associated with the first author (e.g., features numbered 1 until 9 in Table 1) and some document-related factors (e.g., features numbered 22, 24, 26, 28, 29, 49 and 50 in Table 1) were extracted directly from the Web of Science database. Features numbered 23, 25, 33, and 34 in Table 1 are first extracted from WOS and then calculated in Excel software.

In order to extract some features, the HistCite software[3] has been used. To find out the first author's qualitative information (gender, scientific degree, level of education, field of study), the first author's academic page, Google Scholar profile, and otherwise, the first author's profile on social science networks were checked and verified in various ways. Moreover, the rank of the first author's affiliation was determined using the Leiden rankings. The Leiden's ranking uses WOS information for evaluation, like the present research data source.

Self-citations can have a variety of types: e.g., author's self-citation, journal self-citations, or institutional self-citations. On the other, self-citation can be available at citing documents or in cited references (Aksnes 2003a). To calculate authors' self-citations, we considered the similarity between the title of works written by the authors and the title of citing documents or cited references. It should be noted that the WOS database automatically calculates the "Authors' self-citations" number for all of the author's works which seems to be a measure of the author's evaluation; however, we only need the Authors' self-citations for the highly cited article target in order to evaluate the impact of a highly cited article; we obtain this number manually using Excel.

The first author's focus in a specialized field was also obtained by using Gephi Software[4] based on the density of the citation network of the first authors' works (see Small & Griffith 1974)). The title length and abstract length in features numbered 20 and 21 in Table 1 were obtained using R software.

### c) Feature Selection

*Third step*: Generally, filtering, wrapper, and embedded methods are three types of feature selection used in the relevant literature. Two of these, the Lasso (Tibshirani 1996) and Ridge regressions (Marquardt & Snee 1975) are, among the ''embedded methods'', two techniques based on adding a term "as a penalty" to regularize the regression function (Tibshirani 1996). Regularization penalizes models, makes the complex models simpler, and also more stable with smaller regression coefficients while not reducing their predictive power. In other words, Lasso and Ridge regressions use regularization by adding tuning parameters or L1 and L2 penalty terms. In Lasso regression, L1 penalty limits the size of the coefficients. This means that L1 penalty is equal to the absolute value of the magnitude of coefficients. Therefore, some coefficients become zero. On the other hand, L2 penalty, used in Ridge regression, is equal to the square of the

---

magnitude of coefficients. Hence, L2 will yield to the shrinkage of the coefficients but not eliminate any of them (Bühlmann & Van De Geer 2011).

Thus, we stress that by applying a penalty value to the regression model with a large number of variables, these methods, such as the Lasso or Ridge, can reduce the model prediction error on the coefficients of the independent variables in the regression function (Kassambara 2018). On one hand, since a Ridge regression shrinks the value of coefficients no coefficient value reaches zero; thus, all variables are retained in the model. On the other hand, the Lasso regression shrinks coefficients down to zero; whence this type of regression is used to select only the most important variables (features) to be included in the final model.

Recall that traditional procedures such as the Ordinary Least Squares (OLS) regression, the Stepwise regression, and the Partial Least Squares regression are very sensitive to random errors, see e.g. Farahani et al. (2016); that is why various advancements have been proposed in the literature during the past few decades, such as the Ridge regression and the Lasso regression, beside other variants (Muthukrishnan & Rohini 2016). The consideration of these is suggested for future work, maintaining our present study in a coherent (reasonably finite size) frame.

Another feature selection method that is used in our study is the Boruta method, a "wrapper feature selection method". In wrapper methods, a subset of features is used to train a model. Based on the previous model's inference, it is decided to add or eliminate features from the subset (Li et al. 2014; Maldonado et al. 2015). The Boruta algorithm is based on the Random Forest classification algorithm (Kursa & Rudnicki 2010). The method is so used to confirm the final list of features as the most important ones (Kursa & Rudnicki 2010). The method can be implemented in R by using "Boruta package" (Kursa & Rudnicki 2010).

For some completeness, let us mention that several data mining techniques have been reported and discussed, in this journal. Our work differs from such publications with much originality. However, recent papers by Oleinik (2022), Sinclair-Desgagné (2021), Franzosi (2021), Rawat and Sood (2021), Chen et al. (2014) may be mentioned as being of interest to readers for increasing the relevance size of the bibliography toward connected domains.

## 4. Data Analysis

The regression analysis method has been used in many articles related to the present type of research (Figg et al. 2006; Haslam et al. 2008; Paiva et al. 2012; Didegah & Thelwall 2013; Vanclay 2013; Antoniou et al. 2015; Onodera & Yoshikane 2015; Alimoradi et al. 2016; Sohrabi & Iraj 2017; Xie et al. 2019). In our study, because of the continuous dependent variable multicollinearity among the variables, and the high dimensionality of the dataset, that is, there is a large number of independent variables compared to the number of records, we have selected regression regularization methods for our quantitative approach. It has been shown that these methods can be used as a feature selection method by assigning coefficients to each of the features (variables); e.g., see Nie et al. (2010); Tibshirani (2011); Muthukrishnan & Rohini (2016); S. Zhang et al. (2018).

The present data set consisting of 68 observations (the "sample") and 51 features in two categories of document-related and author-related features is presented in Table 1 (including the target feature) where one can find the list of variables, their definitions, and their associated statistical properties according to the type of variables provided by the R software; the *skimr* package (McNamara et al. 2018) was used.

After the initial data cleansing, the next steps prepare the data for analysis. The tasks performed in the preprocessing step are selected based on the data problems and on the technique that is used for the analysis. Accordingly, the problems in the collected data are the existence of missing values and data with zero variance, i.e., a constant value, so that the effect of variable changes could not be examined. The preprocessing tasks related to the techniques are also determined by the fact that the data has to be prepared for correlation and regression analyses. Thus, preprocessing includes the transformation of the distribution of variables to approach the normal distribution for using the Pearson correlation coefficient, normalizing the data in terms of the range of values for correlation and regression analyses, grouping the categorical data that had too many unique values into some main categories, and finally encoding the categorical variables that are needed for correlation and regression analyses. In this regard, let us comment on a few "technical problems," "technique-related requirements," and their solution:

- Missing values: Our strategy in dealing with missing values is to omit the variable if the number of its missing values exceeds 10% of the total number of observations. Thus, "the first author's field of study" with 12 unknown values was eliminated at this stage.
- Zero variance: "Document archived at ResearchGate", "Document archived at Semantic Scholar", and "Document archived at Mendeley" were eliminated due to the zero variance.

Besides, we consider:

- Transformation for skewness: in order to make a correct use of the Pearson correlation coefficient (Pearson 1895), the distribution of high skewness variables was transformed to a normal distribution using Yeo-Johnson (2000) power transformation.
- Normalization: For correlation analysis, as well as Ridge regression, all the values of variables are adjusted to a common scale by subtracting the mean and dividing such a difference by the standard deviation.
- Create dummy variables: categorical variables are converted into binary codes based on the number of their unique values by using a kind of encoding so called one-hot-encoding. Thereafter, one can use the Pearson's correlation coefficient between either quantitative or quantitative variables and logical variables.

All the feature selection steps have been performed using R software and related packages, including *tidyverse* (Wickham 2019), *recipes* (Kuhn & Wickham 2019), and *glmnet* (Friedman et al. 2010).

In order to remove variables that are not correlated with the target (dependent) variable, a correlation analysis was performed after the preprocessing step, through the Pearson correlation coefficient (Pearson 1895).

The list of variables, in a rather insignificant order, the resulting Pearson correlation coefficients, plus other technical informations are given in Table 1. A preprocessing variable number (PPVN) is given in Table 1 for further reference and data comparison.

**Table 1.** Variables list and Pearson correlation analysis results; WOS stems for "Web of Science". PPVN is a preprocessing variable number.

| PPVN | Variables (Related Works) | | | Pearson correlation coefficient | Variable Type (Source of data collection) | Number of missing values | Number of unique values/ frequencies of zero and one | Average (standard deviation) |
|---|---|---|---|---|---|---|---|---|
| 0 | **No. of citations of the document** | | | **Dependent (Target) variable** | **Numerical (WOS)** | **0** | - | **119.41 (97.23)** |
| 1 | **No. of first author's citations** | | | 0.290 | Numerical (WOS) | 0 | - | 1789.06 (1999.39) |
| 2 | **First author's h-index** | | | 0.1698 | | 0 | - | 17.13 (11.11) |
| 3 | **No. of first author's documents** | | | 0.088* | | 0 | - | 74.6 (76.52) |
| 4 | **No. of first author's open-access documents** | | | 0.1446 | | 0 | - | 31.26 (31.82) |
| 5 | **No. of first author's review papers** | | | 0.0911* | | 0 | - | 2.76 (4.33) |
| 6 | **No. of first author's conference papers** | | | 0.0577* | | 0 | - | 14.16 (15.67) |
| 7 | **No. of unique names used by first author in all of her/his documents indexed** | | | 0.0998* | | 0 | - | 4.25 (1.92) |
| 8 | **First author's scientific age**: author's latest working year minus the first working year | | | 0.1449 | | 0 | - | 22.51 (14.7) |
| 9 | **First author's productivity**: The number of author's papers divided by his/her scientific age | | | -0.038* | | 0 | - | 3.44 (2.69) |
| 10 | A | **First author's scientific degree** | Full Professor | 0.0227* | Character (Author's Academic Page/ Social Science Networks) | 0 | 6 | - |
|  | B | | Associate Professor | 0.0889* | | | | |
|  | C | | Assistant Professor | -0.1391 | | | | |
|  | D | | Researcher | 0.038* | | | | |
|  | E | | Lecturer | 0.0227* | | | | |
|  | F | | Student | -0.167 | | | | |
| 11 | **First author's field of study** | | | Removed | | 12 | 6 | - |
| 12 | A | **First author's level of education**: author's last degree | Postdoc | -0.0275* | | 3 | 6 | - |
|  | B | | Pharm.D. | 0.00625* | | | | |
|  | C | | PhD | 0.1304 | | | | |
|  | D | | Master | -0.1931 | | | | |
|  | E | | MD | -0.0609* | | | | |
|  | F | | Dual degree | 0.04198* | | | | |
|  | G | | Other | -0.0228* | | | | |
| 13 | A | **First author's gender** | Female | 0.06345* | | 3 | 2 | - |
|  | B | | Male | -0.0529* | | | | |
|  | C | | Unknown | -0.0228* | | | | |
| 14 | **First author has LinkedIn profile** | | | 0.0157* | Logical (LinkedIn) | 5 | 1s: 46/ 0s: 17 | - |
| 15 | **First author has ResearchGate profile** | | | -0.0925* | Logical (ResearchGate) | 0 | 1s: 47/ 0s: 21 | - |
| 16 | **First author has Google Scholar profile** | | | -0.1638 | Logical (Google Scholar) | 0 | 1s: 44/ 0s: 24 | - |

| PPVN | Variables (Related Works) | Pearson correlation coefficient | Variable Type (Source of data collection) | Number of missing values | Number of unique values/ frequencies of zero and one | Average (standard deviation) |
|---|---|---|---|---|---|---|
| 17 | **First author has Academia profile** | 0.01996* | Logical (Academia) | 0 | 1s: 10/ 0s: 52 | - |
| 18 | **Rank of the first author's affiliation in the Leiden rankings** | -0.00996* | Numerical (Leiden rankings) | 0 | - | 77.49 (136.16) |
| 19 | **First author's focus in a specialized field**: density of citation network of first author's works | 0.1223 | Numerical (Gephi) | 0 | - | 0.02 (0.025) |
| 20 | **Abstract share from paper title**: Frequency of repetition of title words in abstract divided by abstract length (without stop-words) | -0.1580 | Numerical (Excel) | 0 | - | 0.17 (0.087) |
| 21 | **Title share from abstract**: No. of title words in abstract divided by title length (without stop-words) | -0.1337 | | 0 | - | 0.59 (0.14) |
| 22 | **No. of citations by international researchers**: Number of citing documents the target paper authored by researchers with a nationality other than the author of the target article | 0.85585 | Numerical (WOS) | 0 | - | 47.56 (51.39) |
| 23 | **Average age of the references**: Average year of publication of references for each document minus the year of publication of that document | 0.0815* | | 0 | - | 6.99 (5.08) |
| 24 | **No. of references** | 0.11657 | | 0 | - | 53.81 (31.74) |
| 25 | **Average No. of citations of references** | -0.0128* | | 0 | - | 367.1 (387.75) |
| 26 | **Article length**: No. of article pages | -0.0998* | | 0 | - | 7.69 (2.26) |
| 27 | **Authors' self-citations in citing documents**: No. of documents cited in the target article were written by at least one of the authors of the target article | 0.43669 | | 0 | - | 10.71 (12.76) |
| 28 | **Publication year** | -0.8556 | | 0 | - | 2014 (2.61) |
| 29 | **Co-authors**: No. of authors | -0.2368 | | 0 | - | 7.53 (5) |
| 30 | **Journal self-citations in citing documents** | 0.7852 | | 0 | - | 12.29 (13.8) |
| 31 | **Author's self-citations in references**: No. of target article's references which have been written by at least one of the target article's authors | -0.1795 | | 0 | - | 5.25 (5.49) |
| 32 | **Journal self-citations in references**: No. of times the target journal is cited in the references of the article | 0.0813* | | 0 | - | 6.74 (7.15) |
| 33 | **No. of national funding organizations** | -0.1047 | | 0 | - | 1.69 (1.51) |
| 34 | **No. of international funding organizations** | -0.2122 | | 0 | - | 0.21 (0.53) |
| 35 | **No. of collaboration with other organizations** | -0.218 | Numerical (HistCite) | | | 3.82 (2.99) |
| 36 | **No. of co-organization collaboration**: collaboration of authors from the same organization | -0.1629 | | 0 | - | 3.74 (3.40) |

| PPVN | Variables (Related Works) | | Pearson correlation coefficient | Variable Type (Source of data collection) | Number of missing values | Number of unique values/ frequencies of zero and one | Average (standard deviation) |
|---|---|---|---|---|---|---|---|
| 37 | **Title length**: Number of title words without punctuation | | -0.0954* | Numerical (R software) | 0 | - | 13.43 (4.55) |
| 38 | **Abstract length**: Number of abstract characters irrespective of space character and punctuation | | -0.323 | | 0 | - | 205.97 (55.09) |
| 39 | **No. of figures** | | -0.1256 | Numerical (Document texture) | 0 | - | 2.16 (1.79) |
| 40 | **No. of tables** | | -0.16499 | | 0 | - | 2.44 (1.58) |
| 41 | **Title has punctuations** | | 0.01398* | Logical (R software) | 0 | 1s: 53/ 0s: 15 | - |
| 42 | **At least one author from US** | | -0.1622 | Logical (HistCite) | 0 | 1s: 61/ 0s: 7 | - |
| 43 | **Cited in Wikipedia** | | 0.19595 | Logical (Wikipedia) | 0 | 1s: 5/ 0s: 63 | - |
| 44 | **Document has formulae** | | 0.1079 | Logical (Document texture) | 0 | 1s: 7/ 0s: 61 | - |
| 45 | **Document archived at ResearchGate** | | Removed | Logical (ResearchGate) | 0 | 1s: 68 | - |
| 46 | **Document archived at Semantic Scholar** | | Removed | Logical (Semantic Scholar) | 0 | 1s: 68 | - |
| 47 | **Document archived at Mendeley** | | | Logical (Mendeley) | 0 | 1s: 68 | - |
| 48 | **Title contains a question** | | -0.0087* | Logical (Document Title) | 0 | 1s: 3/ 0s: 65 | - |
| 49 | **Open-access paper**: Is the paper open access? | | 0.3794 | Logical (WOS) | 0 | 1s: 65/ 0s: 3 | - |
| 50 | A | **Document type** | Review | 0.1354 | Character (WOS) | 0 | 2 | - |
| | B | | Article | -0.1354 | | | | |

The * next to the value of the correlation coefficient of some variables indicates the weak correlation of these variables with the number of citations. Notice also that some skewness of the data can be observed through the confidence interval width with respect to the mean of the data distribution.

It can be rightfully claimed that variables with a correlation coefficient of less than 0.1 have no relationship or have an ignorable relationship. Accordingly, 28 variables specified in Table 1 can be omitted because of the weak correlation with the target variable (i.e., variables marked with an asterisk next to their correlation coefficient were removed). As a result, 32 independent variables (including dummy variables) were used in the next step.

Due to this ''large'' number of data points, it is an information step to look for the rank-size relationship of the Pearson Correlation coefficients. It is shown on Fig. 1 that (removing the 5 apparently meaningless, at first imagined to be relevant, independent variables, see Table 1), the Pearson correlation coefficients are distributed into 3 regimes, or sets, indeed. This observation points to the most important external factors and allows us to develop the study accordingly.

*Determining the coefficients of variables using Ridge and Lasso methods:*

The coefficient of each independent variable in the regression model is next determined using both Ridge and Lasso methods. Figures 2 and 3 show the mean square error for Ridge and Lasso cases, respectively, as a function of the logarithm of $\lambda$, where $\lambda$ is the value that determines the reduction rate of the coefficients of the variables to minimize the cross-validation error (Kassambara 2018). This error has some variance/standard error, depicted by the grey whiskers to every red point. The numbers on the top of the graph give the number of non-zero regression coefficients in our model (thus, the number of included features). From left to right along the x-axis, with increasing $\lambda$, in Fig. 3, (since there are fewer variables in the model); indeed, we recall that the penalty for inclusion of features is weighted more heavily in the Lasso method.

The coefficients of the 32 variables, which are assessed as relevant at this stage, obtained by the Ridge method and sorted from the highest coefficient to the lowest one are given in Table 2. Moreover, the corresponding coefficient sign is given in view of indicating either the positive or negative effect on the target variable.

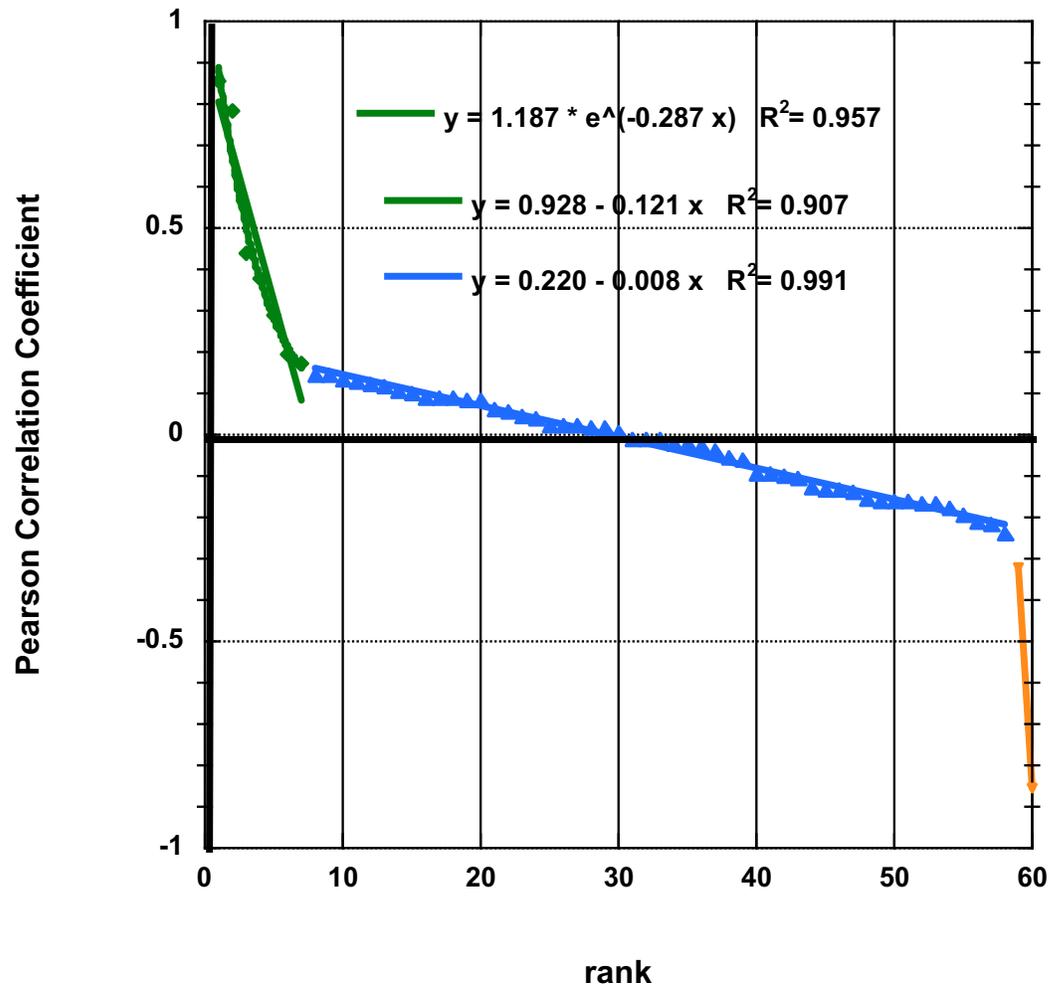

**Fig. 1** Pearson correlation coefficient ranking pointing to 3 different (linear or quasi linear, i.e. weakly exponentially decreasing) regimes, whence to external factors importance.

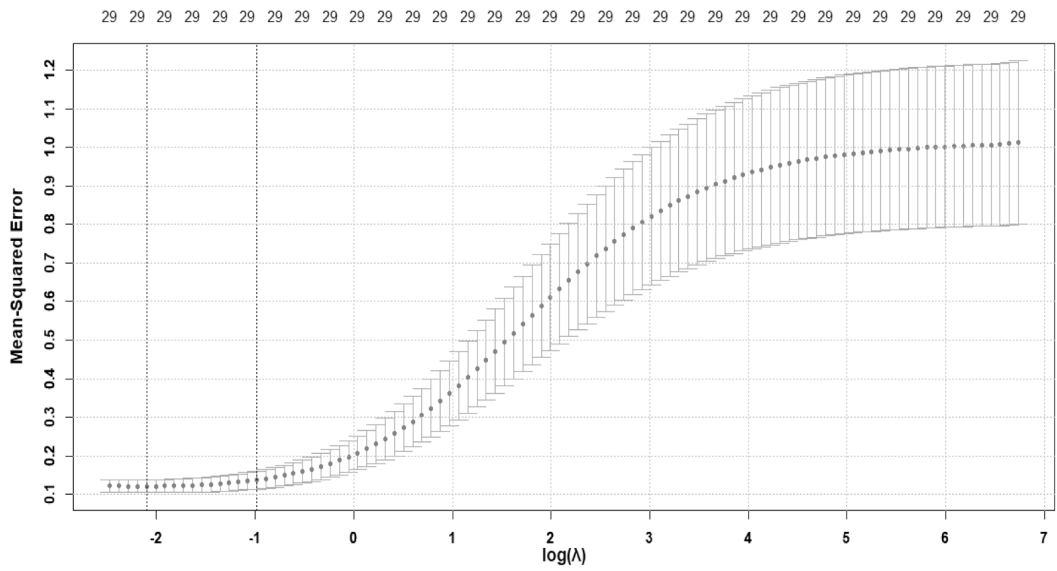

**Fig. 2** Mean squared error as a function of the logarithm of λ in the Ridge method

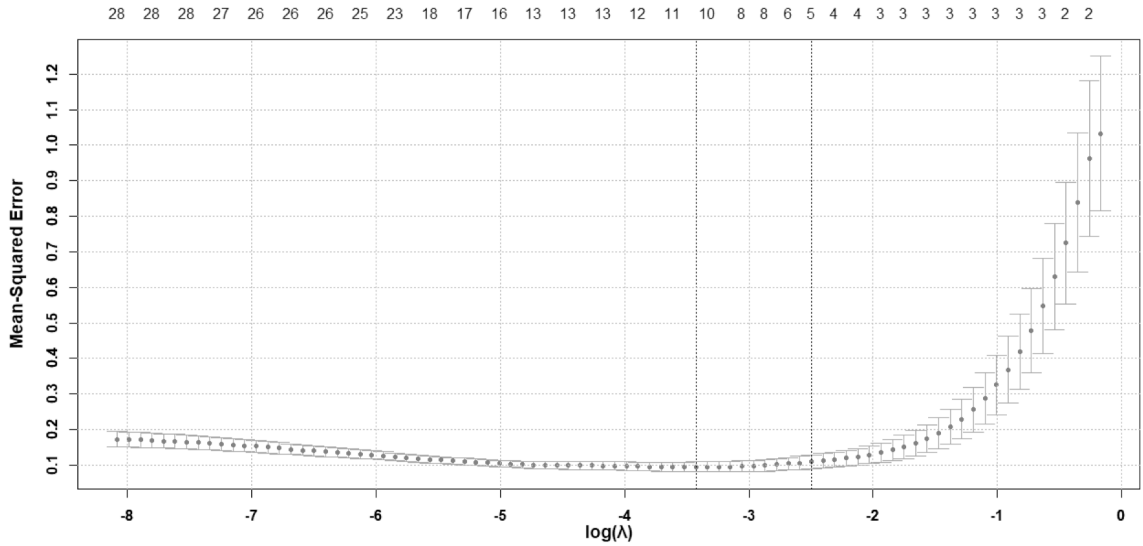

**Fig. 3** Mean squared error as a function of the logarithm of λ in the Lasso method

Out of 32 variables in Table 2, 19 variables do not show a significantly positive coefficient indicating a positive factor for increasing the number of citations.

Then, the 13 variables, only those with positive coefficients, are used for applying the Ridge method for the second time. In other words, the Ridge regression applied to these 13 variables, so the related coefficients are changed. Table 3 shows these variables in decreasing "importance order" based on the coefficients so obtained.

**Table 2.** Variable regression coefficients using the Ridge method.

PPVN : preprocessing variable number (see Table 1).

| Rank | PPVN | Variable name | Coefficients based on the Ridge method | | Rank | PPVN | Variable name | Coefficients based on the Ridge method | |
|---|---|---|---|---|---|---|---|---|---|
| | | | Value | Sign | | | | Value | Sign |
| 1 | 22 | No. of citations by international researchers | 0.34997 | Positive | 17 | 24 | No. of references | 0.0227 | Positive |
| 2 | 30 | Journal self-citations in the citing documents | 0.2338 | Positive | 18 | 8 | First author's scientific age | 0.0221 | Positive |
| 3 | 28 | Publication year | 0.2262 | Negative | 19 | 33 | No. of national funding organizations | 0.0194 | Negative |
| 4 | 27 | Authors' self-citations in the citing documents | 0.1167 | Positive | 20 | 29 | Co-authors (No. of authors) | 0.0155 | Negative |
| 5 | 34 | No. of international funding organizations | 0.05532 | Negative | 21 | 20 | Abstract share from paper title | 0.0096 | Negative |
| 6 | 1 | No. of first author's citations in WOS | 0.0438 | Positive | 22 | 43 | Cited in Wikipedia | 0.00897 | Positive |
| 7 | 19 | First author's focus in a specialized field | 0.0401 | Negative | 23 | 40 | No. of tables | 0.0081 | Negative |
| 8 | 12 | Level of Education – PhD | 0.0387 | Positive | 24 | 21 | Title share from abstract | 0.0075 | Negative |
| 9 | 38 | Abstract length | 0.0383 | Negative | 25 | 10c | Scientific degree – assistant professor | 0.007 | Negative |
| 10 | 31 | Authors' self-citations in references | 0.0376 | Negative | 26 | 10f | Scientific degree – student | 0.00696 | Negative |
| 11 | 49 | Open-access paper | 0.0311 | Positive | 27 | 36 | No. of co-organization collaboration | 0.0045 | Negative |
| 12 | 44 | Document has formulae | 0.0311 | Positive | 28 | 50b | Document type – Article | 0.0044 | Negative |
| 13 | 35 | No. of collaboration with other organizations | 0.0293 | Negative | 29 | 50a | Document type – Review | 0.00438 | Positive |
| 14 | 39 | No. of figures | 0.0291 | Negative | 30 | 4 | No. of first author's open-access documents in WOS | 0.00397 | Positive |
| 15 | 16 | First author has Google Scholar profile | 0.0268 | Negative | 31 | 42 | At least one author from US | 0.0012 | Positive |
| 16 | 12d | Level of Education – Master | 0.02299 | Negative | 32 | 2 | First author's h-index in WOS | 0.0001 | Negative |

**Table 3.** Coefficients obtained for variables with a positive effect on the citation number using the Lasso method; the ranking is in the decreasing order of the coefficient value. PPVN : preprocessing variable number (as in Table 1).

| Rank | PPVN | Variable name | Coefficient value based on the Lasso regression method |
|------|------|---------------|--------------------------------------------------------|
| 1 | 22 | No. of citations by international researchers | 0.6147 |
| 2 | 30 | Journal self-citations in citing documents | 0.3651 |
| 3 | 27 | Authors' self-citations in citing documents | 0.0860 |
| 4 | 8 | First author's scientific age | 0.0368 |

It should be noted that using the Lasso method, Table 3, only four variables, i.e., "Number of citations by international researchers", "Number of journal's self-citation in citing documents", "Number of authors' self-citation in citing documents", and "First author's scientific age" are of significant importance and have a finite coefficient.

Finally, as mentioned in the Methodology section, the Boruta method is used to confirm the results achieved by Ridge and Lasso methods. Table 4 presents the outcome of the Boruta method.

In Table 4, NormHits is the number of hits normalized to the number of importance source runs, and Decision represents whether the variable can be considered important, i.e., "Confirmed," or has a very low importance score and can be neglected, i.e., "Rejected." In Table 4, only the confirmed variables are listed.

**Table 4.** The most important features with a marked effect on the citation number using the Boruta method. PPVN : preprocessing variable number (see Table 1).

| PPVN | Variable name | Mean-Importance | NormHits | Decision |
|------|---------------|-----------------|----------|----------|
| 28 | Publication year | 20.66076 | 1.00000 | |
| 22 | No. of citations by international researchers | 20.58589 | 1.00000 | |
| 30 | Journal self-citations in citing documents | 14.31397 | 1.00000 | |
| 27 | Authors' self-citations in citing documents | 5.38936 | 1.00000 | Confirmed |
| 49 | Open-access paper | 3.78165 | 0.86869 | |
| 1 | No. of first author's citations in WOS | 1.84994 | 0.42424 | |

The result of feature selection by the Boruta method confirms that "No. of citations by international researchers", "Journal self-citations in citing documents", "Authors' self-citations in citing documents", "Open-access paper", and "No. of first author's citations in WOS" are the most important features in terms of predictive power in the ''Number of citations'' of highly cited papers. Notice that "First author's scientific age" is not found as so relevant after the Boruta filtering. On the other hand, "Publication year" is also recognized as an important variable by the Ridge method but negatively affected the number of citations. So, this variable was eliminated for further discussion. Clearly, by using a combination of these methods and results, a better insight into important features can be achieved.

## 5. Discussion

*Number of citations by international researchers*

By calculating the number of articles that cited a target article authored by researchers with a different nationality from the nationalities of the target article authors, we find that highly cited papers are more frequently cited by international researchers, - within our sample. This finding is consistent with the results of Aksnes (2003b). Since the number of citations is the presumably best indicator of the impact/quality of a research (Kostoff 2007), receiving citations from the international community can bring more transnational influence and global credibility to the authors than citations from the national community.

*Journal's self-citations and authors' self-citations*

The self-citation rate is one of the evaluation criteria for journals indexed on the Web of Science. The share of self-citations in highly cited papers was found insignificant by Noorhidawati et al. (2017) and Aksnes (2003b). The results of our study show that there is a negative and weak correlation between the presence of the target papers' authors and the target journal in the references of the target papers; however, there is a positive and significant relationship between the number of citing documents published in the target journal and the number of citations of the target papers. Thus, the authors' self-citations number is also a significantly positive factor. Fowler and Aksnes (2007) have also shown that "the more one cites oneself, the more one is cited by other scholars". It might be recalled that self-citations might indicate creativity, in particular when an author is apparently changing his/her field of research (Hellsten et al. 2006, 2007; Ausloos et al. 2008).

It should be said about the journal's self-citation in citing documents that since each journal often focuses on a specific field, it would be more likely that articles published in the same field and citing each other will be published in a similar journal. This can be even truer for journals with a high impact factor. Therefore, it cannot be claimed that self-citations have led to a high citation of these papers. Vanclay (2013), who has examined this variable in his article and observed its high impact, states that this fact does not indicate the encouragement toward self-citations, but rather the "desirability of publishing within a journal where a conversation is taking place". The same is true of the authors' self-citation in citing documents feature. Some researchers follow their new research works based on the results of their previous works, so they cite their previous works, and when their previous works have a high intrinsic quality, the probability of citing them increases (Ausloos et al. 2008).

*Number of authors, First author's scientific age and Total Number of first author's citations in WOS*

The author's eminence can predict the impact of a document (Haslam et al. 2008). Accordingly, one may easily assume that many researchers check the authors' scientific works' validity and reputation when considering whether to use (to read) and cite a scientific paper. Obviously, the first author's role in an article is usually not equal to that of the other authors. The first author usually has the most significant role in producing a research work (Gaeta 1999), - when the alphabetical order is not imposed for whatever reason. Therefore, if the first author has more experience and has higher citations in his research profile, then this may indicate his/her contribution credibility.

Moreover, the number of authors has a positive effect on increasing the citation rate (Figg et al. 2006; Falagas et al. 2013; Fox et al. 2016). Since the number of authors reflects the wide scope of scholarly collaboration (Dorta-González & Santana-Jiménez 2019), more collaborations would be likely to be reflected into more research ideas.

Also, self-archiving can affect the number of citations; this effect can be confirmed by increasing the number of self-archiving through more contributors to an article. Nevertheless, other reports contradict that claim; for example, Onodera & Yoshikane (2015) showed that the number of authors and their achievements, - out of 13 other external factors, had the least impact on the citation rate. Ho (2014) also stated that there is no significant relationship between the number of authors in a highly cited papers and its number of citations. In this regard, the present study leads to conclude that co-authoring does not positively affect the number of citations. This is somewhat unexpected, and might be a journal effect.

*Open-access and Cited in Wikipedia*

Open-access to the research work is likely expected to increase the number of its reading. In so doing, the "open release" of a research output extensively raises concerned people's awareness, which leads to more citations (Ale Ebrahim et al. 2013). However, citing an article may depend on the degree of its visibility rather than the merit of the article (Marashi et al. 2013). For example, when a research work is included in Wikipedia, the number of citations increases over time (Marashi et al. 2013). The results of our study also confirm that "open-access" is an important feature that increases the citations of research work, but to be "cited in Wikipedia" is relatively less important.

*Document type*

The document type also affects the number of citations e.g. as observed by Onodera & Yoshikane (2015). For example, some studies have shown that review papers are more cited than other types of papers (Alimoradi et al. 2016; Aksnes 2003b), which may be due to the larger number of references in the review articles (Stevens et al. 2019). Indeed, these articles are a combination of ideas and findings of previous research in a field (Lei and Sun 2020). Some studies have even suggested that researchers seeking to attract citations may write a review paper instead of an article (Vanclay 2013). As our analysis shows, when the highly cited paper is a review paper, its citation rate is higher than the original paper. However, this feature was not confirmed by any feature selection methods used in this research, whence can be considered as an effective feature, - though with less statistical confidence than others.

*Other factors*

Like most previous studies, our Ridge method study confirms that a large number of references positively influences the number of citations (Antoniou et al. 2015; Onodera & Yoshikane 2015; Fox et al. 2016; Zhang & Guan 2017; Dorta-González & Santana-Jiménez 2019; Stevens et al. 2019). The number of references can indicate the author's knowledge of previous research (Dorta-González & Santana-Jiménez 2019). It should be noted that this feature is considered relatively less important by other feature selection methods. Besides, our findings on the title length are consistent with the research of Letchford et al. (2015) and Alimoradi et al. (2016). However, due to modern retrieval technologies, nowadays, one article with a long title can get more citations because it includes more words, whence it would be more likely to be retrieved by search engines (Guo et al. 2018). Concerning the question titles or using other punctuation marks in the title our findings are in line with research by Jamali & Nikzad (2011), Paiva et al. (2012), and Stevens et al. (2019): articles with question titles are more frequently downloaded but less frequently cited (Jamali & Nikzad, 2011). Interestingly, Van Wesel et al. (2014) pointed out that the effect of these factors depends on the disciplines: social sciences papers have more complex titles than ''hard science'' papers. Notice that little or no correlation is found here between the gender of the first author and the number of citations, - as in Stevens et al. (2019) or Haslam et al. (2008).

## 6. Conclusions

*Back to the research question: "what are the most important features of highly cited papers?"*

According to the results obtained in Table 2 (Ridge method), Table 3 (Lasso method), and Table 4 (Boruta method), it can be observed that "No. of citations by international researchers", "Journal self-citations in citing documents", "Authors' self-citations in citing documents" are recognized as important features by all three methods we have used. Moreover, "First author's scientific age", "Open-access paper", and "No. of first author's citations in WOS" have to be considered important by two methods only. Therefore, these six features are the final features that this research can offer as the most important or relevant ones.

Notice that the findings of the present study do not mean that one can claim that the citation rate of highly cited papers has been pre-manipulated up to now, - as it is sometimes found (Herteliu et al. 2017). We agree with Wang et al. (2019) that individuals cannot manipulate these factors to turn their article into an ESI highly cited paper because "*the possibility of becoming ESI highly cited papers depends not only on their own quality, but also on the quality of other papers and the number of papers published each year*".

Accordingly, although the research findings are challenging and can be discussed, - as always there are limitations, so far (to the best of our knowledge), we emphasize that these 3 feature selection methods (exclusively Ridge, Lasso, and Boruta) have not been used to identify the most important features of highly cited articles elsewhere. Thus, the methods so used can be a new way for scientometrics and altmetrics research in citation prediction studies, beside others. Filters can be imagined in order to improve the quality of work to be published. By removing a few limitations of the present study, more (and ''better'') results can be even obtained through the here above proposed, discussed, and illustrated feature selection methods.

In conclusion, we emphasize the performance resulting from such algorithms. However, we do not advise authors to seek to increase the citations of their articles by manipulating the identified performance features. Indeed, ethical rules regarding these characteristics must be strictly obeyed.

## Acknowledgements

Great thanks to reviewers.

## Funding

Marcel Ausloos is partially supported by a grant of the Romanian National Authority for Scientific Research and Innovation, CNDS-UEFISCDI, project number PN-III-P4-ID-PCCF-2016-0084.

## Conflicts of interest

The authors claim to have no conflict of interest.

## Contributions of authors

All authors contributed to the study conception and design to the best of their own skills and expertise. All authors commented on previous versions of the manuscript. All authors read and approved the final manuscript.